\documentclass[aps,twocolumn,showkeys,showpacs,preprintnumbers,prd,superscriptaddress,nofootinbib,10pt]{revtex4-2}
\usepackage{graphicx,epsf,bm,amsmath,amsfonts,amssymb,epstopdf,natbib,hyperref,color,verbatim,multirow,bm,appendix}
\hypersetup{colorlinks=true,urlcolor=blue,citecolor=blue,linkcolor=blue,menucolor=blue,anchorcolor=blue,filecolor=blue}
\DeclareUnicodeCharacter{2212}{-}
\DeclareUnicodeCharacter{202F}{ }

\begin{document}

\title{Status of early dark energy after DESI: the role of $\Omega_m$ and $r_s H_0$}

\author{Jun-Qian Jiang}
\email{jiangjunqian21@mails.ucas.ac.cn}
\email{jiangjq2000@gmail.com}
\affiliation{School of Physical Sciences, University of Chinese Academy of Sciences, No.19(A) Yuquan Road, Beijing 100049, China}

\begin{abstract}
\noindent The EDE model is one of the promising solutions to the long-standing Hubble tension.
This paper investigates the status of several EDE models in light of recent BAO observations from the Dark Energy Spectroscopic Instrument (DESI) and their implications for resolving the Hubble tension.
The DESI Y1 BAO results deviate from the CMB and Type Ia supernova (SNeIa) observations in their constraints on the matter density $\Omega_m$ and the product of the sound horizon and the Hubble constant $r_s H_0$.
Meanwhile, these EDE models happen to tend towards this deviation.
Therefore, in this work, it is found that DESI Y1 BAO results strengthen the preference for EDE models and help to obtain a higher $H_0$.
Even considering the Pantheon+ observations for SNeIa, which have an opposite tendency, DESI still dominates the preference for EDE.
This was unforeseen in past SDSS BAO measurements and therefore emphasizes the role of BAO and SNeIa measurements in Hubble tension.
\end{abstract}

\maketitle
\newpage

\section{Introduction}
\label{sec:introduction}

The Hubble tension~\cite{Verde:2019ivm,DiValentino:2021izs,Perivolaropoulos:2021jda,Schoneberg:2021qvd,Shah:2021onj,Abdalla:2022yfr,DiValentino:2022fjm,Hu:2023jqc,Verde:2023lmm} refers to the persistent discrepancy between the Hubble constant ($H_0$) values derived from early-Universe observations, such as the Cosmic Microwave Background (CMB), and those obtained from local-Universe measurements, including Cepheid-calibrated type Ia supernovae. 
For example, Planck 2018 reported $H_0 = (67.4\pm 0.5)$ km/s/Mpc~\cite{Planck:2018vyg} base on the CMB measuments while the SH0ES team reported $H_0 = (73.04 \pm 1.04)$ km/s/Mpc~\cite{Riess:2021jrx} based on Cepheid-calibrated type Ia supernovae. 
This discrepancy ($4 \sim 6 \sigma$~\cite{Verde:2019ivm}) has become a significant challenge in cosmology, prompting investigations into potential solutions.

Various explanations have been proposed to address this tension, ranging from systematic errors in measurements to extensions of the $\Lambda$CDM model.
One of the promising beyond $\Lambda$CDM models is Early Dark Energy (EDE, see e.g. Refs.\cite{Karwal:2016vyq,Poulin:2018cxd,Kaloper:2019lpl,Agrawal:2019lmo,Lin:2019qug,Smith:2019ihp,Niedermann:2019olb,Sakstein:2019fmf,Ye:2020btb,Gogoi:2020qif,Braglia:2020bym,Lin:2020jcb,Odintsov:2020qzd,Seto:2021xua,Nojiri:2021dze,Karwal:2021vpk,Rezazadeh:2022lsf,Poulin:2023lkg,Odintsov:2023cli,Sohail:2024oki,Rezazadeh:2022lsf}),
a hypothetical form of energy that would have been present in the early Universe, accelerating its expansion rate $H(z)$ prior to recombination.
As a result, the sound horizon $r_s = \int c_s(z)/H(z) dz$ is reduced.
On the other hand, the constraint on $H_0$ from CMB measurements mainly comes from precise measurements of the acoustic peak angle separation $\theta_s$:
\footnote{In this work, the difference between $r_s^*$ defined by the optical depth (relevant for CMB) and $r_s^\text{drag}$ defined by the baryon drag epoch (relevant for BAO) is ignored because the models considered here will not significantly change the difference between them.
All values of $r_s$ shown in this paper are defined by the baryon drag epoch.}
\begin{equation} \label{eq:theta_s}
    \theta_s = \frac{r_s}{D_A} \propto r_s H_0 \, .
\end{equation}
As EDE does not significantly change the late cosmic expansion history, \autoref{eq:theta_s} maintains its form, whereas $H_0$ deduced from it has a higher value for EDE.
In this way, the $H_0$ inferred from CMB observations gets closer to the results inferred from local observations such as SH0ES.
In addition, the EDE model also has the potential to relieve other observational tensions, such as the abundance of high redshift galaxies observed by JWST~\cite{Forconi:2023hsj,Liu:2024yan,Shen:2024hpx}.
It has been found that current JWST observations support the EDE model even without considering Hubble tension~\cite{Jiang:2024tll}.

Some observations of the late universe, such as Baryon Acoustic Oscillations (BAO) and uncalibrated Type Ia supernovae (SNeIa), can be used to extract information on the background universe.
This information is used to constrain the expansion history of the late Universe, and these constraints can be applied in (cosmological) model-independent ways, which limits the possibility of resolving the Hubble tension by modifying the late Universe alone~\cite{Bernal:2016gxb,Addison:2017fdm,Lemos:2018smw,Aylor:2018drw,Schoneberg:2019wmt,Knox:2019rjx,Arendse:2019hev,Efstathiou:2021ocp,Krishnan:2021dyb,Cai:2021weh,Keeley:2022ojz,Jiang:2024xnu} and reinforces the necessity for EDE-like modifications to the early Universe.
At the same time, for the EDE model, if we assume that the late Universe remains $\Lambda$CDM, this information can provide constraints on the density of matter $\Omega_m$ today (see also its importance in sound horizon independent constraints~\cite{Kable:2024mgl}).
In addition, BAO also provides constraints on $r_s h$, where $h = H_0 / (100$ km/s/Mpc).
They can all be used to strengthen the constraints on the corresponding parameters in the EDE model, which further translates into the strengthening of other parameters through degeneration between parameters.

The presence of EDE suppresses the growth of the gravitational potential in the corresponding time, which would alter the CMB physics that depends on it, such as the early ISW effect, the potential envelope, and lensing.
Therefore, when fitting the CMB, physical matter density $\omega_m = \Omega_m h^2$ has to be simultaneously increased to enhance perturbation growth as EDE increases $H_0$~\cite{Ye:2020oix,Vagnozzi:2021gjh,Jiang:2024nha}.
This is similar to the relationship between $\omega_m$ and $H_0$ after fixing $\Omega_m$ by BAO and SNeIa measurements.
However, this is actually a coincidence.
These observations are not necessarily fully compatible with the EDE model parameters of the CMB constraints, and they are in fact an important independent constraint.
On the other hand, the constraints of these observations on $\Omega_m$ are also important in leading to the enhancement of late-time matter perturbations (parameterized by $\sigma_8$ or $S_8$) in these early-time solutions of Hubble tension, such as EDE~\cite{Pedrotti:2024kpn,Poulin:2024ken}.
This enhancement strengthens the tension with observations of large-scale structures, which is one of the main challenges of the EDE model~\cite{Hill:2020osr,Ivanov:2020ril,DAmico:2020ods}.

There are some mild inconsistencies in these recent BAO and SNeIa measurements of these quantities.
Recent BAO observations favor a smaller $\Omega_m$ relative to Planck 2018 results.
For example, SDSS reported $\Omega_m = 0.299 \pm 0.016$~\cite{eBOSS:2020yzd} and DESI Y1 reported $0.295 \pm 0.015$~\cite{DESI:2024mwx}.
In contrast, recent SNeIa observations, such as Pantheon+~\cite{Brout:2022vxf}, favor a larger $\Omega_m$ relative to Planck 2018 results.
See e.g.~\cite{Colgain:2024xqj,Colgain:2024mtg,Colgain:2024ksa} for discussions on the $\Omega_m$ tension.
Meanwhile, DESI Y1 BAO prefers a large $r_s h = 101.8 \pm 1.3$ Mpc relative to Planck 2018 results, which is a $1.9 \sigma$ discrepancy between them~\cite{DESI:2024mwx}.
As explained earlier, these constraints not only can tighten the constraints on the parameters of the EDE model, but the inconsistency between them may also lead to some different constraints.

In this work, I investigate different EDE models and update their state after DESI Y1 BAO data.
With these results, I show how DESI Y1's constraints on $\Omega_m$ and $r_s h$ can enhance preferences for EDE and its potential to resolve Hubble tension, as well as attenuate the tension with large-scale structures.

The EDE models investigated in this work are briefly summarised in \autoref{sec:ede}.
The datasets and analysis methods used are described in \autoref{sec:data}.
The results are presented in \autoref{sec:results} and discussed there.
Finally, I conclude in \autoref{sec:conclusions}.

\section{Early dark energy models}
\label{sec:ede}

Here I investigate five typical EDE models that have different approaches to energy decay and different behavior at perturbation level.

\subsection{axion-like EDE}

The axion-like EDE model~\cite{Poulin:2018cxd,Smith:2019ihp} has a potential similar to the axion with higher-order instanton corrections, which is in practice difficult in theory though, we can also treat it as a toy model:
\begin{equation}
    V(\phi) = m^2 f^2 [1-\cos(\phi/f)]^n, \quad \theta = \phi/f \in [0, 2\pi],
\end{equation}
where $m$ is the mass and $f$ is the axion decay constant.
At first, the field is held in a position $\Theta_\text{ini}$ by Hubble friction, and thus has dark energy-like properties.
Then as the background expansion rate decreases, the field rolls towards the bottom of the potential and loses energy in oscillation.
The decay rate of the EDE energy, described by the equation of state parameter $w$, is controlled by the index of the potential: $w=(n-1)/(n+1)$.
As the fit to the data supports that $n=3$ (i.e. $w=1/2$) is the closest integer to the observation, investigations of axion-like EDE models typically fix it, and hence I similarly pick the $n=3$ axion-like EDE model.
I perform a phenomenological investigation here, i.e. I convert two of the field theory parameters $m, f$ into phenomenological parameters: the redshift at which the EDE field starts to roll $z_c$ and the energy fraction of EDE at that time $f_\text{EDE}$.
There is one remaining parameter $\Theta_\text{ini}$, which controls the shape of the potential and affects the speed of sound $c_s^2$.
I use the publicly available modified \texttt{CLASS} code\footnote{\url{https://github.com/mwt5345/class_ede_v3.2.0}}.

\subsection{$\phi^4$ AdS-EDE}

AdS-EDE~\cite{Ye:2020btb} is a strongly theory-motivated class of models.
The anti-de Sitter (AdS) vacuum is ubiquitous in string theory~\cite{Bousso:2000xa,Danielsson:2009ff}, and it naturally provides a way to make the EDE energy decay rapidly.
Here I investigate the original $\phi^4$ AdS-EDE, which has a potential:
\begin{equation}
    V(\phi)=\left\{\begin{array}{ll}
    V_{0}\left(\dfrac{\phi}{M_{\text{Pl}}}\right)^{4}-V_{\text{AdS}} & ,\quad \dfrac{\phi}{M_{\text{Pl}}}<\left(\dfrac{V_{\text{AdS}}}{V_{0}}\right)^{1 / 4} \\
    0 &, \quad\dfrac{\phi}{M_{\text{Pl}}}>\left(\dfrac{V_\text{AdS}}{V_{0}}\right)^{1 / 4}
    \end{array}\right.
\end{equation}
where $M_{\text{Pl}}$ is the reduced Planck mass and $V_{\text{AdS}}$
is the depth of AdS phase.
After the potential unfreezes from the Hubble friction, it rolls towards the bottom of the potential through an AdS phase before eventually entering the $V=0$ phase.
During the AdS phase, it decays energy extremely fast with $w>1$ and maintains $c_s^2=1$.
The magnitude of the potential at the final stage can also be modified as today's cosmological constant.
The field theory parameters of the AdS-EDE are usually converted into three phenomenological parameters: $\alpha_{\text{AdS}} = V_\text{AdS} / \left(\rho_\text{m}(z_c)+\rho_\text{r}(z_c)\right)$ describes the relative depth of the AdS potential, which is fixed to the value chosen by Ref.\cite{Ye:2020btb}.
$z_c$ is the redshift of $\partial_{\phi}^2 V(\phi_c) = 9 H_c^{2}$, which is the approximate moment when the potential begins to roll.
And $f_\text{EDE}$ is the energy fraction of EDE at that time.
One of the concerns about the AdS-EDE model is “AdS bound”, which means that sometimes the field does not have enough energy to climb out of the AdS phase into the third phase.
This means that the theoretical prior does not actually fill the space described by the phenomenological parameters.
\footnote{The $\phi^4$ AdS-EDE model investigated here is only a simple toy AdS-EDE model.
There can be other realizations of AdS-EDE model with different kinds of potential, which can avoid the issue of incomplete prior~\cite{Ye:2020oix}.}
Nevertheless, here I use it as an example to examine the case when (early universe-based) $H_0$ is brought to achieve the SH0ES result.
The modified \texttt{CLASS} code is publicly available\footnote{\url{https://github.com/genye00/class_multiscf}}.

\subsection{canonical ADE}

Acoustic dark energy (ADE)~\cite{Lin:2019qug} is a phenomenological model using a perfect fluid with the equation of state:
\begin{equation}
    1 + w(a) = \frac{1+w_f}{[1+(a_c/a)^{3(1+w_f)/p}]^p} \, .
\end{equation}
ADE has $w\sim-1$ at $a<a_c$, after which it transforms to $w\sim w_f$. The rapidity of this transition is determined by $p$, which is fixed to $1/2$.
I use the same notation $f_\text{EDE}$ to denote the energy fraction of ADE at $a_c$.
Moreover, the behavior of ADE at the perturbation level is determined by a constant effective sound speed $c_s^2$ in its rest frame.
In this work, I fix $c_s^2 = w_f=1$, which corresponds to a canonical scalar and has been investigated in~\cite{Lin:2019qug}.

\subsection{Cold NEDE}

New Early Dark Energy (NEDE)~\cite{Niedermann:2019olb,Niedermann:2020dwg} achieves the energy decay of EDE through a phase transition.
In the case of the cold NEDE considered here, this phase transition is introduced by the trigger field $\phi$:
\begin{equation}
    V(\psi, \phi) = \frac{\lambda}{4} \psi^4 + \frac{\beta}{2} M^2 \psi^2 - \frac{1}{3} \alpha M \psi^3 + \frac{1}{2} m^2 \phi^2 + \frac{1}{2} \bar{\lambda} \phi^2 \psi ^2 \, .
\end{equation}
In the beginning, the NEDE field $\psi$ is in a false vacuum.
When the expansion rate $H$ of the universe decreases such that $H \lesssim m$, $\phi$ rolls down, degrading the potential barrier between the false vacuum and the true vacuum, which triggers a first-order phase transition of tunneling into the true vacuum.
NEDE is realized as an effective fluid with different equations of state before and after the decay time:
\begin{equation}
    w_\text{NEDE} = \begin{cases}
        -1, & \text{for } z>z_\text{decay} \\
        \omega_\text{NEDE}, & \text{for } z \leq z_\text{decay}
    \end{cases} \, .
\end{equation}
The decay time is determined by the trigger field $\phi$ with an initial value $\phi_\text{ini}$ and a trigger condition parameter $H_\text{decay}/m$.
In this work, I follow~\cite{Niedermann:2020dwg} and fix $\phi=10^{-4} M_\text{pl}$ and $H_\text{decay}/m=0.2$.
The fraction of NEDE at the time of decay is denoted as the usual $f_\text{EDE}$.
In addition to these, in order to describe the perturbations of the NEDE fluid, we need the effective sound speed in the fluid rest-frame $c_s^2$ and the viscosity parameter $c_\text{vis}^2$.
Here, I fix the latter to 0 and the former to be equal to the adiabatic sound speed.
In summary, I vary three parameters $\{ f_\text{EDE}, z_\text{decay}, \omega_\text{NEDE}\}$.
I use the publicly available modified \texttt{CLASS} code\footnote{\url{github.com/NEDE-Cosmo/TriggerCLASS}}.

\subsection{$\phi^4$ Rock ‘n' Roll}

The Rock ‘n' Roll model~\cite{Agrawal:2019lmo} is a simple EDE model using a scalar field with potential $V \propto \phi^{2n}$.
Similar to the axion-like EDE model, its energy decays through the oscillatory behavior after thawing.
Actually, it corresponds to the limit $\Theta_\text{ini} \rightarrow 0$ of the axion-like EDE model and also has $w = (n-1)/(n+1)$ during the oscillatory phase.
In this work, I fix $n=2$, which is the best-fit value found in~\cite{Agrawal:2019lmo}.
Therefore, the $\phi^4$ Rock ‘n' Roll can be described by only two phenomenology parameters: the thawing redshift $z_c$ when $\partial_{\phi}^2 V(\phi_c) = 9 H_c^{2}$ is satisfied, and the EDE energy fraction $f_\text{EDE}$ at $z_c$.

\section{Datasets and methodology}
\label{sec:data}

\begin{table}[htb!]
\centering
\begin{tabular}{|c|c|} \hline \hline
Parameters & Prior \\ \hline
$\log_{10}(1+z_c)$ & [7.5, 0] \\
$f_\text{EDE}$     & [0, 0.3] \\
$\Theta_\text{ini}$ (axion-like EDE) & [0, 3.14] \\
$\omega_\text{NEDE}$ (Cold NEDE) & [1/3, 1] \\ \hline
$H_0$              & [20, 100] \\
$n_s$              & [0.8, 1.2] \\
$\omega_b = \Omega_b h^2$ & [0.005, 0.1] \\
$\omega_c = \Omega_c h^2$ & [0.05, 0.99] \\
$\log(10^{10} A_\text{s})$ & [1.61, 3.91] \\
$\tau_\text{reio}$ & [0.01, 0.8] \\
\hline \hline
\end{tabular}
\caption{The prior ranges for cosmological parameters used in this work.}
\label{tab:prior}
\end{table}

The primary observation at the low $H_0$ end of Hubble tension is the CMB.
Therefore, CMB observations are included in all the datasets analyzed.
I use the latest available Planck data, which consists of:
the Planck 2018 (PR3) \texttt{Commander} likelihood~\cite{Planck:2019nip} for the low-$\ell$ TT spectrum, the PR4 \texttt{lollipop} likelihood~\cite{Hamimeche:2008ai,Mangilli:2015xya,Tristram:2020wbi} for the low-$\ell$ EE spectrum and the PR4 \texttt{CamSpec} likelihood~\cite{Rosenberg:2022sdy} for the high-$\ell$ TTTEEE spectrum.
They are denoted as “PR4" in this work.
The DESI Y1 (2024) BAO~\cite{DESI:2024mwx} measurements, which were obtained from the analysis of galaxy, quasar, and Lyman-$\alpha$ forest tracers in the redshift range $0.1 < z < 4.2$, are then included in the dataset.
To investigate the constraints from SNeIa observations, measurements from Pantheon+ (PantheonPlus)~\cite{Brout:2022vxf} for the SNeIa light curve in the redshift range $z=0.01$ to 2.26 are also used.

In addition, to compare the DESI results with previous results, I also considered the BAO measurements of the SDSS, which included 6dF~\cite{Beutler:2011hx}, SDSS DR7 main Galaxy sample~\cite{Ross:2014qpa}, SDSS DR12~\cite{BOSS:2016wmc} and SDSS DR16~\cite{eBOSS:2020yzd}.
I used the final BAO-only results present in \cite{eBOSS:2020yzd}.
Furthermore, the inclusion of SH0ES data~\cite{Riess:2021jrx} is also investigated.

I use \texttt{Cobaya}~\cite{Torrado:2020dgo} for Monte Carlo Markov Chain (MCMC) analysis.
The prior ranges of cosmological parameters employed in this work are listed in \autoref{tab:prior}.
All the chains are converged to satisfy the Gelman-Rubin parameter~\cite{Gelman:1992zz} $R-1 < 0.01$.
The prior volume effect may bias marginalized parameter posterior distributions in Bayesian analysis.
Its impact is significant when $f_\text{EDE}$ is low.
Therefore, the best-fit values of parameters are also provided since they are not subject to the prior volume effect.
I've found that the default method of finding the best-fit value in \texttt{Cobaya} is not robust in some cases, i.e. it doesn't converge to the true best-fit point.
In contrast, the best fits in this work are searched by \texttt{prospect}~\cite{Holm:2023uwa}, which minimizes $\chi^2$ by simulated annealing.

\section{Results and discussion}
\label{sec:results}

\begin{figure*}[htb!]
    \centering
    \includegraphics[width=\linewidth]{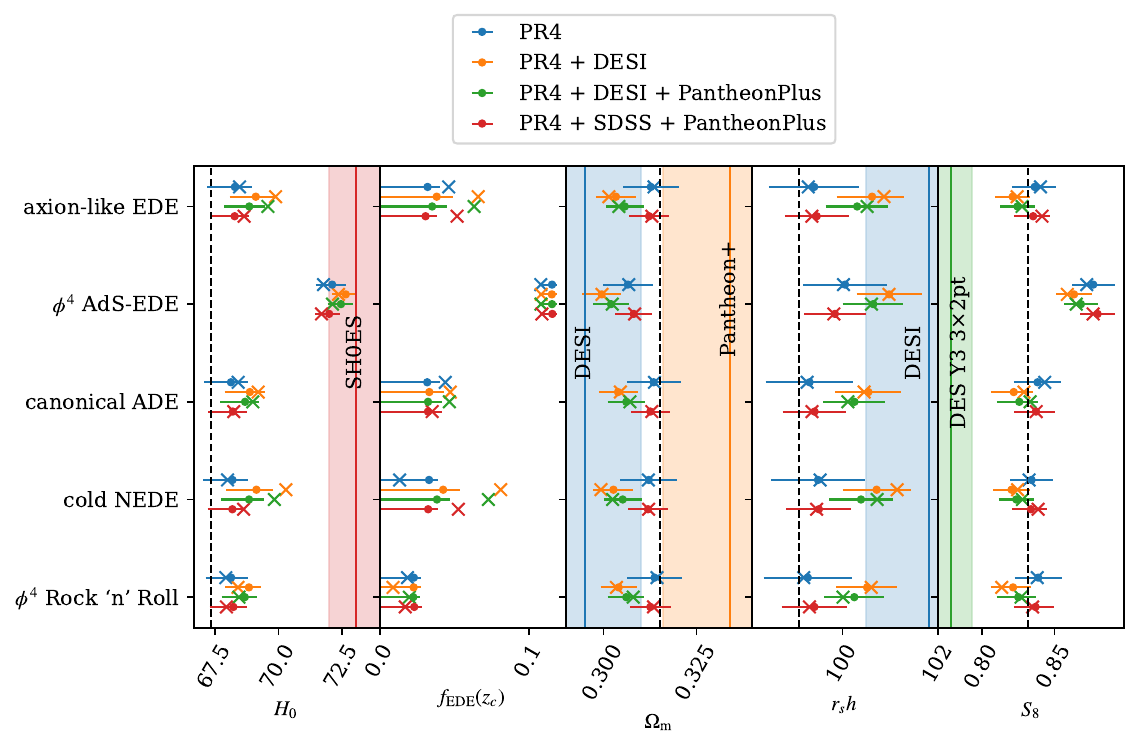}
    \caption{Posterior distributions of the parameters when using the different BAO and SNeIa datasets with the PR4 dataset, where dots indicate means and error bars indicate 68\% confidence intervals. The best-fit values are indicated by crosses.
    The colored bands and solid vertical lines indicate the $1\sigma$ constraints on the parameters from the corresponding observations.
    In addition, the black dashed lines indicate the mean values of the Planck 2018 baseline result.}
    \label{fig:PR4}
\end{figure*}

\begin{figure*}[htb!]
    \centering
    \includegraphics[width=\linewidth]{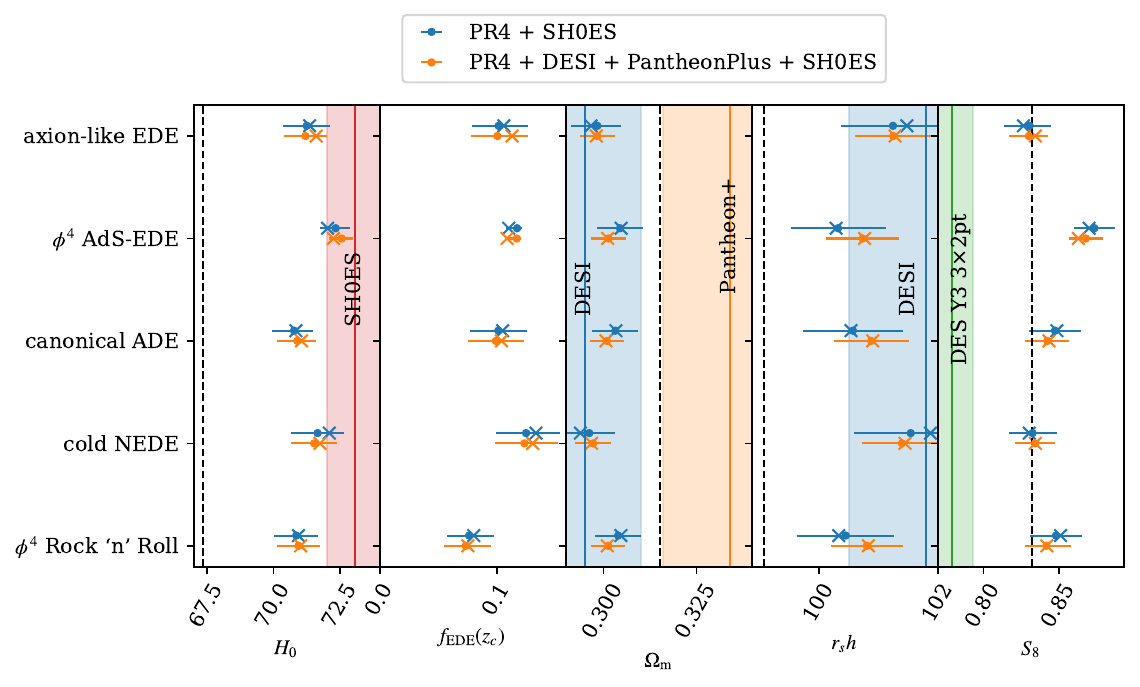}
    \caption{Posterior distributions of the parameters when using the different BAO and SNeIa datasets with the PR4 + SH0ES dataset, where dots indicate means and error bars indicate 68\% confidence intervals. The best-fit values are indicated by crosses.
    The colored bands and solid vertical lines indicate the $1\sigma$ constraints on the parameters from the corresponding observations.
    In addition, the black dashed lines indicate the mean values of the Planck 2018 baseline result.}
    \label{fig:PR4+SH0ES}
\end{figure*}

\begin{figure*}[htb!]
    \centering
    \includegraphics[width=\linewidth]{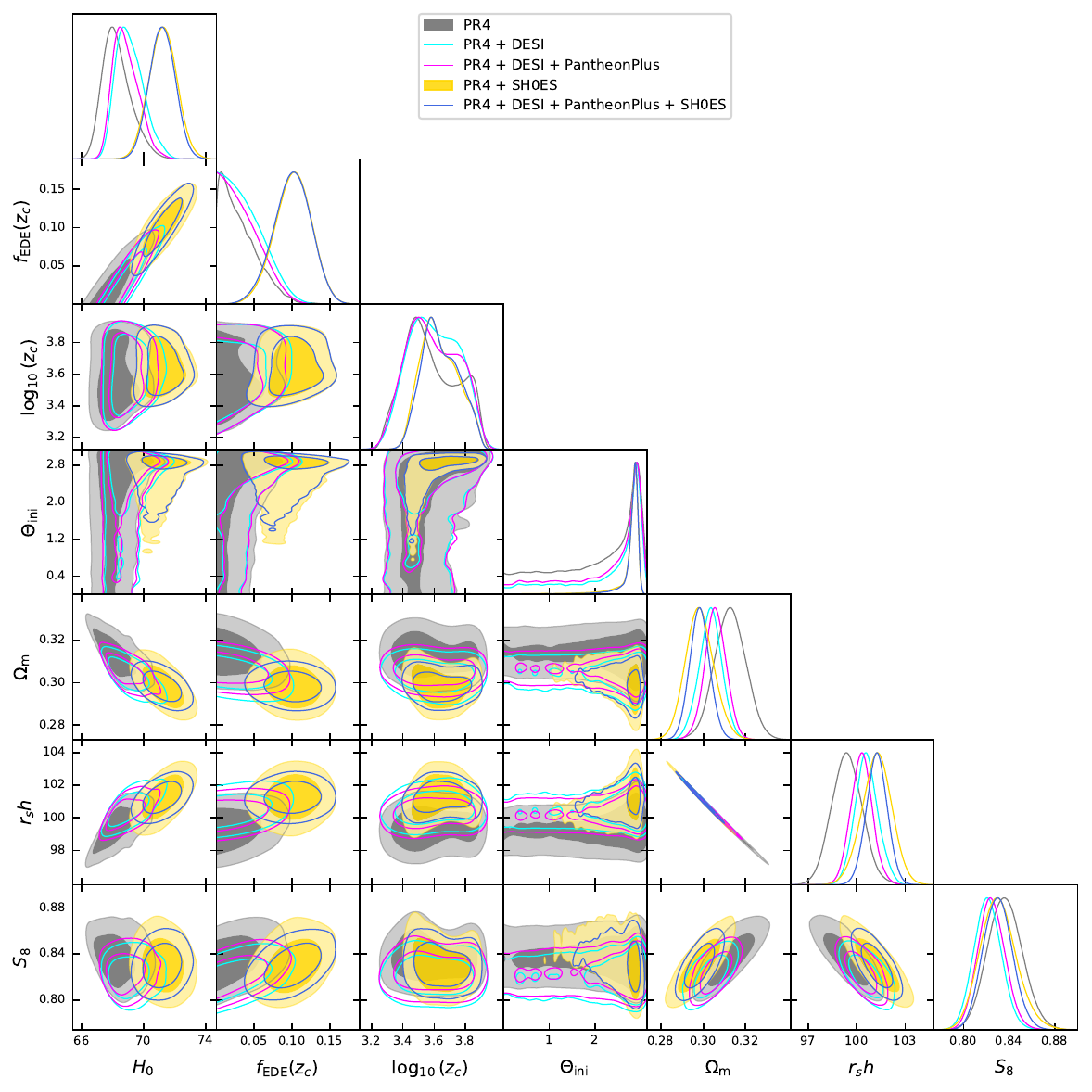}
    \caption{Posterior distributions of the parameters of the axion-like EDE model under different data sets (contour lines indicate 68\% and 95\% confidence intervals).}
    \label{fig:axionEDE}
\end{figure*}

\begin{figure*}[htb!]
    \centering
    \includegraphics[width=\linewidth]{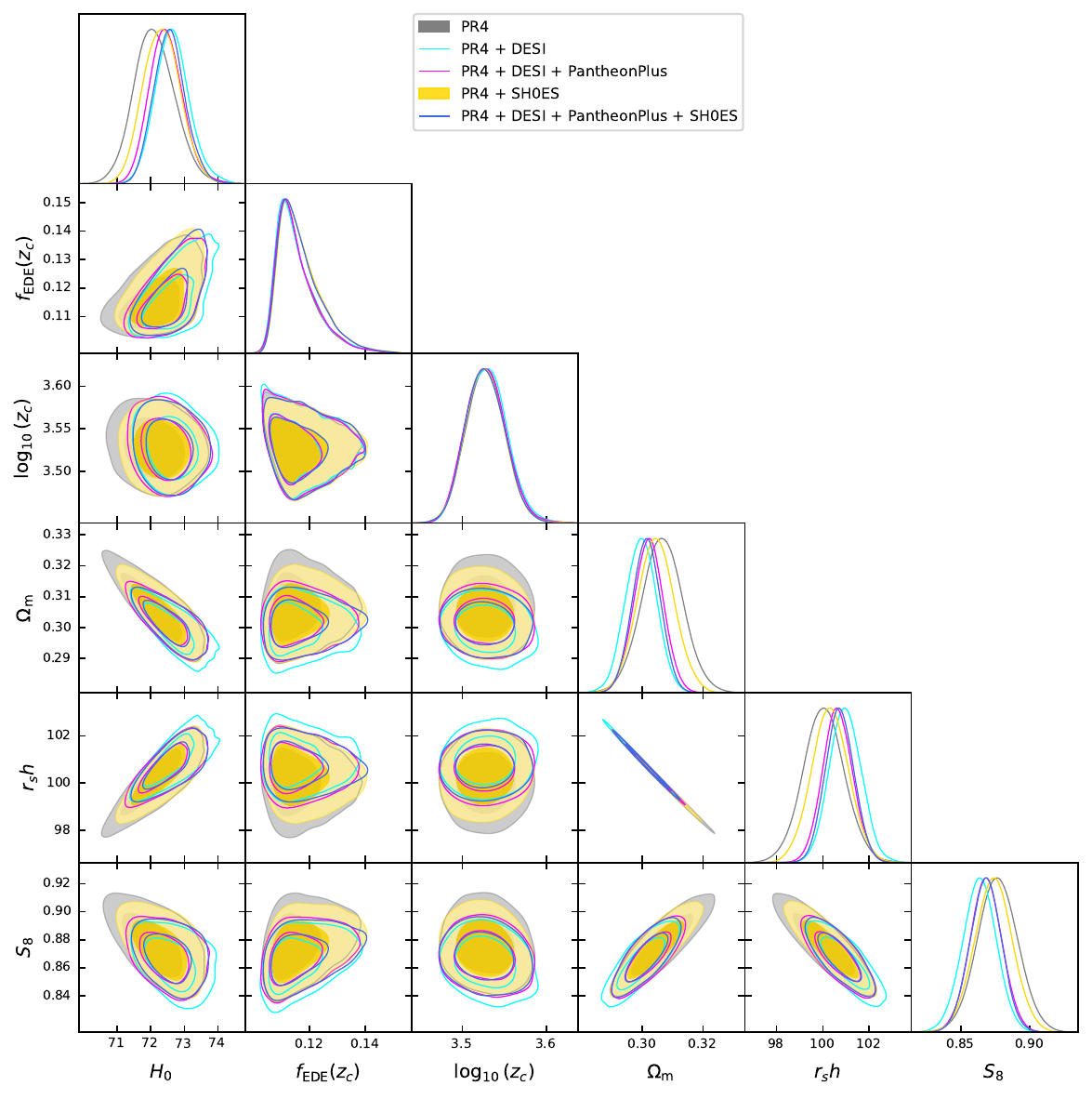}
    \caption{Posterior distributions of the parameters of the AdS-EDE model under different data sets (contour lines indicate 68\% and 95\% confidence intervals).}
    \label{fig:AdSEDE_default}
\end{figure*}

\begin{figure*}[htb!]
    \centering
    \includegraphics[width=\linewidth]{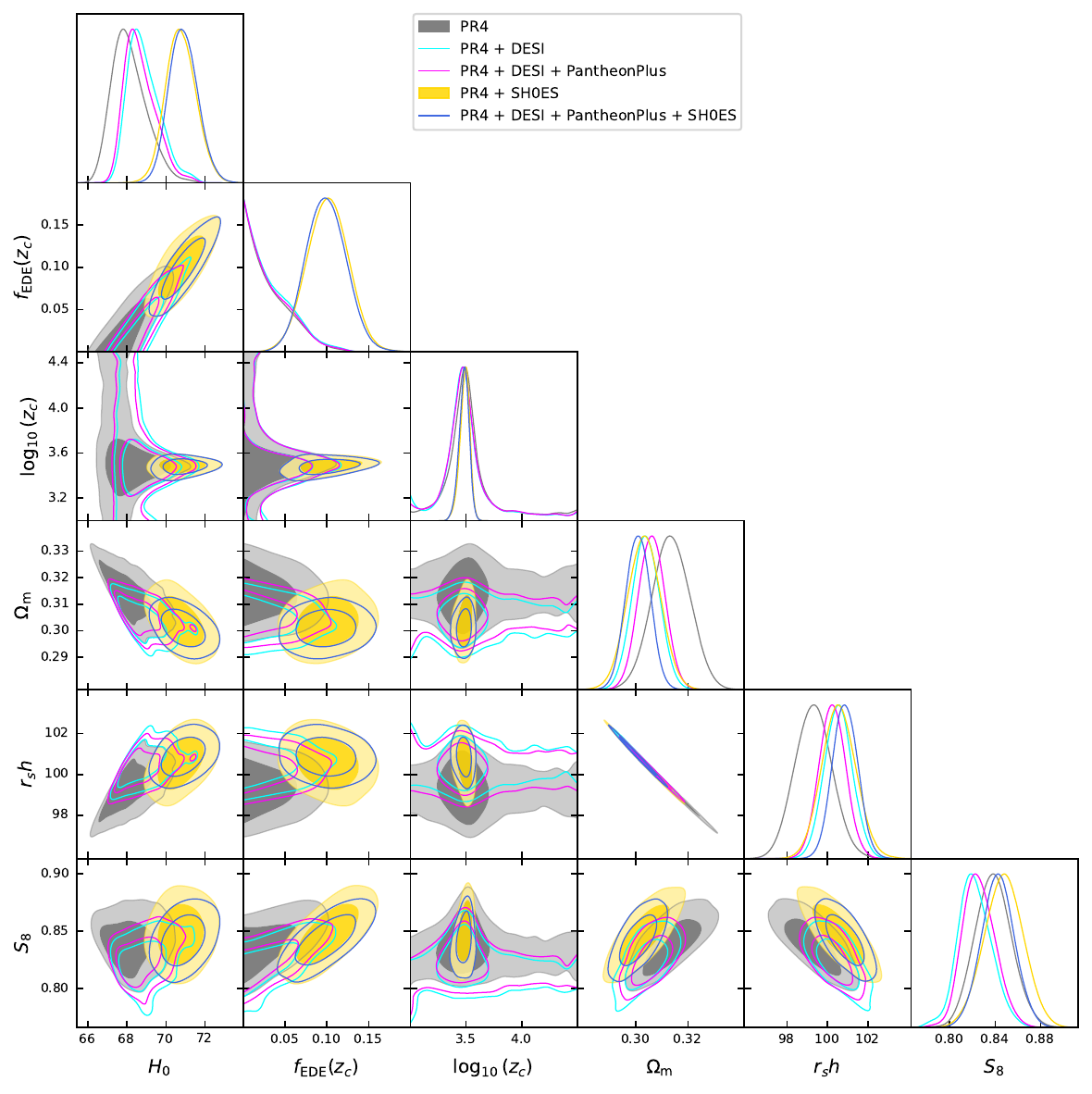}
    \caption{Posterior distributions of the parameters of the canonical ADE model under different data sets (contour lines indicate 68\% and 95\% confidence intervals).}
    \label{fig:cADE}
\end{figure*}

\begin{figure*}[htb!]
    \centering
    \includegraphics[width=\linewidth]{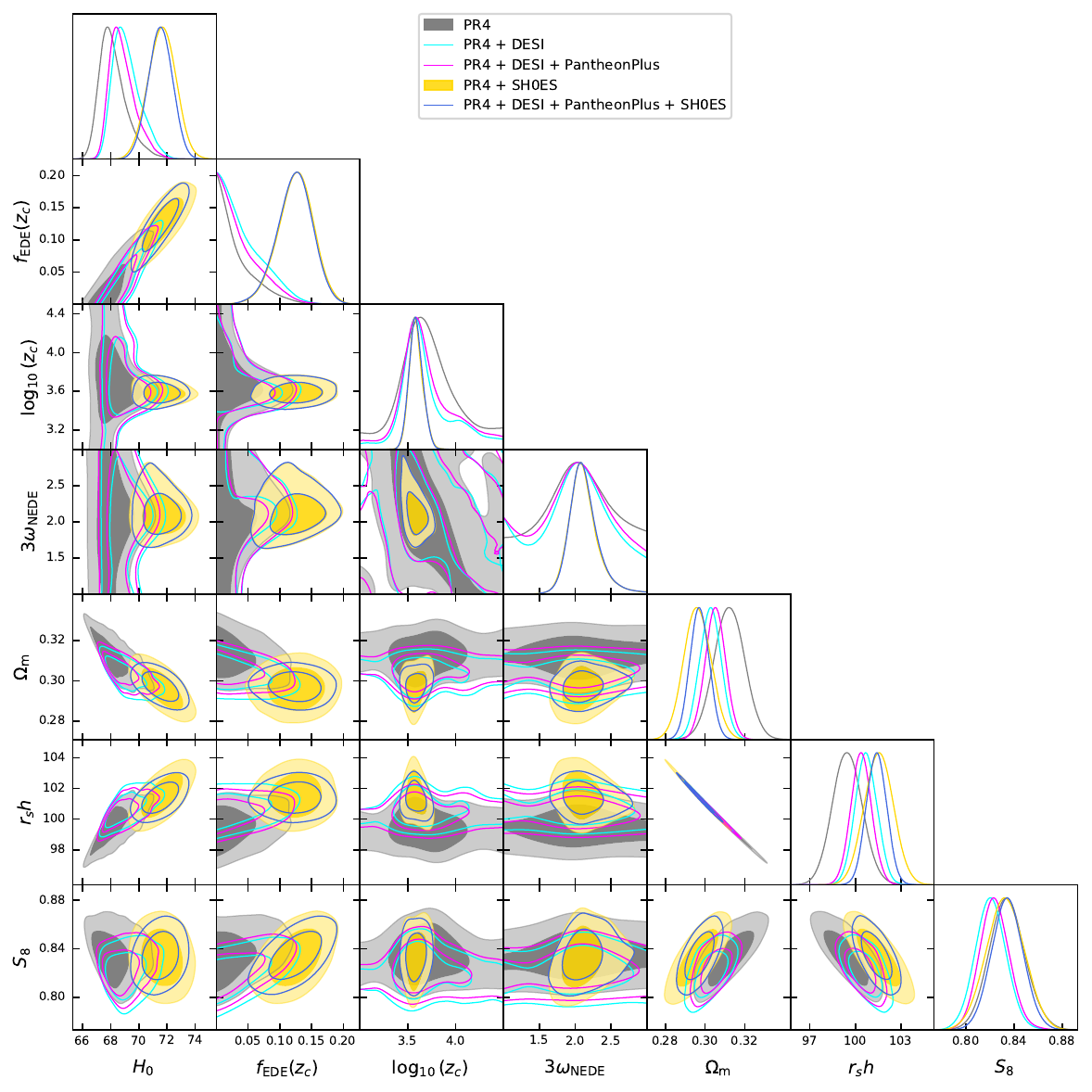}
    \caption{Posterior distributions of the parameters of the cold New EDE model under different data sets (contour lines indicate 68\% and 95\% confidence intervals).}
    \label{fig:NEDE}
\end{figure*}

\begin{figure*}[htb!]
    \centering
    \includegraphics[width=\linewidth]{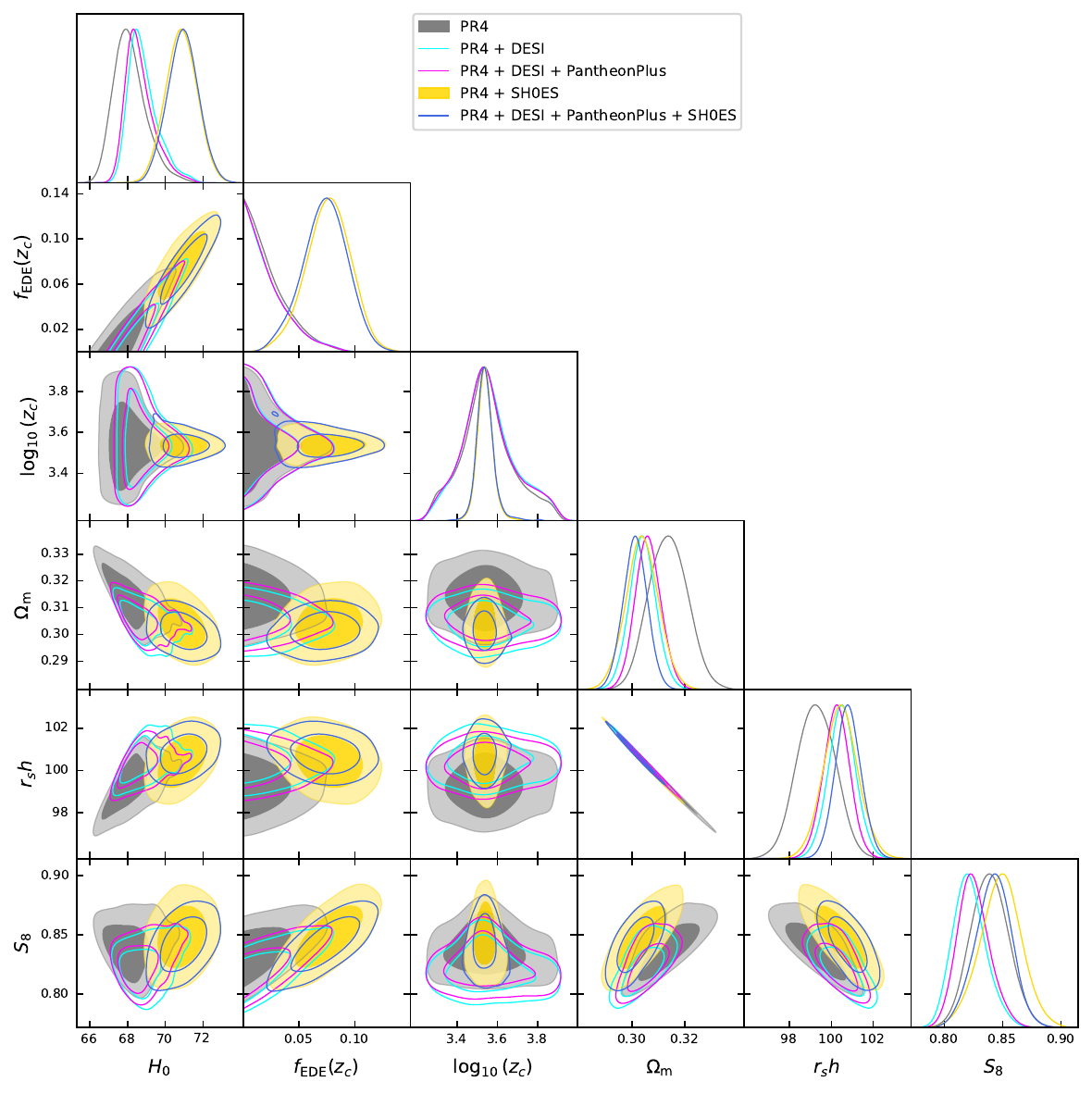}
    \caption{Posterior distributions of the parameters of the $\phi^4$ rock ‘n' roll EDE model under different data sets (contour lines indicate 68\% and 95\% confidence intervals).}
    \label{fig:phi2n_default}
\end{figure*}

The main results of this work are summarized in \autoref{fig:PR4}, which lists the constraints on some of the cosmological parameters with and without BAO/SNeIa measurements when the SH0ES results are not included.
The marginalized posterior distributions of cosmological parameters and their best-fit values for the different datasets are listed in Table \ref{tab:PR4},\ref{tab:PR4_DESI},\ref{tab:PR4_DESI_PP}, respectively.
In addition, in \autoref{fig:PR4+SH0ES} and Table \ref{tab:PR4_SH0ES},\ref{tab:PR4_SH0ES_DESI_PP}, the posterior distribution of the parameters with the addition of SH0ES data is presented.
For each model, the 2D posterior distributions of parameters with different datasets are also shown in Figure \ref{fig:axionEDE},\ref{fig:AdSEDE_default},\ref{fig:cADE},\ref{fig:NEDE},\ref{fig:phi2n_default} respectively.

Firstly, for PR4 alone, all EDE models except $\phi^4$ AdS-EDE are compatible with $f_\text{EDE}=0$
\footnote{It should be noted that different EDE models have slightly different definitions for $z_c$ and $f_\text{EDE}$, and therefore direct comparisons between them should not be made.}
as well as the Planck 2018 baseline $H_0$ within $1\sigma$.
It should be mentioned again that although the $\phi^4$ AdS-EDE model leads to a high $f_\text{EDE}$ and $H_0$ in this case, the theoretical prior of this model does not fully cover the phenomenological parameter space described in \autoref{tab:prior}.
Here I use it only as an example of raising CMB-based $H_0$ to SH0ES results and inspect the shift of other parameters.
The posterior distribution of $\Omega_m$ in the $\phi^4$ AdS-EDE model is lower than the Planck 2018 baseline result in $\gtrsim 1\sigma$.
Meanwhile, the posterior distribution of $r_s h$ is higher than the Planck 2018 baseline result in $\gtrsim 1\sigma$.
When SH0ES data is included, all the EDE models obtain significantly non-zero $f_\text{EDE}$, and all the $H_0$ posterior distributions are raised to $\gtrsim 70$ km/s/Mpc in differing extents.
Moreover, the same trend can be found for $\Omega_m$ and $r_s h$, but the deviation relative to the Planck 2018 baseline results is more obvious due to their larger EDE fractions.
It can also be found there are slight relations between $\Omega_m$ ($r_s h$) and $f_\text{EDE}$ in the 2D posterior distributions, in particular for the axion-like EDE model and NEDE model.
This verifies that EDE models do not guarantee that $\Omega_m$ will be consistent with the results of $\Lambda$CDM, nor do they guarantee that they will be consistent with the results of relevant measurements.
In contrast, these EDE models exhibit deviations in the same direction relative to the $\Lambda$CDM, at least for the models investigated here.
The behavior of $r_s h$ then follows $\Omega_m$, as for a $\Lambda$CDM late universe:
\begin{equation}
    \theta_s = \frac{r_s}{D_A} \propto \frac{r_s h}{\int_0^{z_\text{CMB}} dz/\sqrt{\Omega_m (1+z)^3 + (1-\Omega_m)}} \, ,
\end{equation}
where we can find an inverse relation between $r_s h$ and $\Omega_m$ when $\theta_s$ is fixed by the CMB observations.
This relationship can be clearly found in Figure \ref{fig:axionEDE},\ref{fig:AdSEDE_default},\ref{fig:cADE},\ref{fig:NEDE},\ref{fig:phi2n_default}.
Besides, for axion-like EDE and canonical ADE, their $f_\text{EDE}$ best-fit values are outside the $1\sigma$ range of the corresponding marginalized posterior distributions, suggesting that the prior volume effect may bias Bayesian inference from the frequentist point of view.

The constraints on $\Omega_m$ and $r_s h$ from the DESI Y1 BAO measurements are shown in blue bands in \autoref{fig:PR4}.
Both of them deviate from the Planck 2018 baseline results with $> 1 \sigma$.
Interestingly, they deviate in the same direction as the previously mentioned EDE model deviation relative to the Planck 2018 baseline results.
After DESI observations are added to the PR4 analysis, the posterior distributions of $f_\text{EDE}$ for axion-like EDE, canonical ADE, and cold NEDE all have their upper bounds slightly raised, although they are still compatible with $f_\text{EDE}=0$ at $1\sigma$.
The posterior distribution of $f_\text{EDE}$ for $\phi^4$ Rock ‘n’ Roll model shrinks very slightly with the addition of DESI.
For both axion-like EDE and NEDE, their best-fit $f_\text{EDE}$ are significantly shifted toward high value as they are free from the prior volume effect.
In particular, I notice that the best-fit $f_\text{EDE}$ for NEDE jumps from below the mean to above the mean after adding DESI.
Profile likelihood analysis using \texttt{prospect} suggests that this is because NEDE has two similar best-fit values for $f_\text{EDE}$ (or $H_0$), with the one with lower $f_\text{EDE}$ having the lower $\chi^2$ in the case of PR4.
This confirms the findings in Appendix A of \cite{Chatrchyan:2024xjj}.
\footnote{The combination of PR4 data used in \cite{Chatrchyan:2024xjj} is slightly different from here.}
For all EDE models, $H_0$ is raised, regardless of the Bayesian posterior distributions or the frequentist best-fit values.
This is a consequence of the double effect of higher (or equal) $f_\text{EDE}$ and higher $r_s h$.
Despite this, they (except for $\phi^4$ AdS-EDE) still fail to agree with the SH0ES results.
In addition to this, their $S_8$ becomes lower due to lower $\Omega_m$, although there are still tensions with e.g. the DES Y3 results~\cite{DES:2021wwk}.

Recent SNeIa observations report constraints opposite to BAO for $\Omega_m$, such as Pantheon+, which is shown as an orange band in \autoref{fig:PR4}.
I performed a joint analysis with DESI and PR4.
As shown in \autoref{fig:PR4}, it slightly pulls back $\Omega_m$ toward the Planck 2018 baseline result, which results in slight decreases in $f_\text{EDE}$ and $H_0$ as well as slight increases in $S_8$.
Nevertheless, the shift of the parameters with respect to PR4 is dominated by DESI.

The past BAO measurements from SDSS provided similar parameter uncertainties to the DESI Y1 results.
However, compared to DESI it favors a larger $\Omega_m$ and smaller $r_s h$, i.e. more compatible with the Planck 2018 baseline results (see e.g. Figure 2 in \cite{DESI:2024mwx}).
The results of replacing DESI measurements with SDSS are shown as red error bars in \autoref{fig:PR4}.
The specific values are not listed in this paper for brevity.
In this case, almost all results go back to the PR4 case.
However, there is still some very slight preference for EDE.
For example, the best-fit $f_\text{EDE}$ for NEDE jumps from below the mean to above the mean again.
This can be understood because SDSS still has a lower $\Omega_m$ and higher $r_s h$ than Planck, though not as much as DESI.

Finally, I investigate the inclusion of DESI and Pantheon+ on top of the PR4+SH0ES dataset, which is compared in \autoref{fig:PR4+SH0ES}.
In this case, although $\Omega_m$ and $r_s h$ are slightly shifted by DESI and Pantheon+, the constraints for $H_0$ and $f_\text{EDE}$ are dominated by the SH0ES data and therefore remain almost unchanged.
Besides, the shifts in $n_s$ and $H_0$ listed in Table \ref{tab:PR4_SH0ES},\ref{tab:PR4_SH0ES_DESI_PP} validate the previous finding~\cite{Ye:2021nej,Jiang:2022uyg,Cruz:2022oqk,Jiang:2022qlj,Jiang:2023bsz,Peng:2023bik,Wang:2024tjd,Wang:2024dka} that reaching $H_0 \sim 73$ km/s/Mpc would simultaneously imply that $n_s \simeq 1$ holds for these latest datasets as well, which could have a significant impact on the inflation model (e.g. Refs.~\cite{Kallosh:2022ggf,Ye:2022efx}).

\begin{table*}[htb!]
\fontsize{6}{0}
    \centering
\begin{tabular} {| l | c| c| c| c| c|}
\hline\hline
 & axion-like EDE & $\phi^4$ AdS-EDE & canonical ADE & cold NEDE & $\phi^4$ Rock ‘n’ Roll\\
\hline
$f_\mathrm{EDE}(z_c)       $ & $< 0.0786(0.0461)          $ & $0.1157(0.1082)^{+0.0033}_{-0.0086}$ & $< 0.0820(0.0438)          $ & $< 0.0941(0.0131)          $ & $< 0.0596(0.0184)          $\\
$\log_{10}(z_c)            $ & $3.58(3.879)^{+0.28}_{-0.21}$ & $3.528(3.5256)^{+0.022}_{-0.025}$ & $3.548(3.493)^{+0.089}_{-0.23}$ & $3.71(4.158)^{+0.23}_{-0.32}$ & $3.55(3.539)^{+0.11}_{-0.14}$\\
$\Theta_\mathrm{ini}       $ & $- \ (2.99)              $ &                              &                              &                              &                             \\
$3 \omega_\mathrm{NEDE}    $ &                              &                              &                              & $2.03(1.00)^{+0.62}_{-0.51}$ &                             \\
\hline
$H_0                       $ & $68.29(68.47)^{+0.68}_{-1.1}$ & $72.11(71.77)^{+0.56}_{-0.65}$ & $68.13(68.42)^{+0.68}_{-1.1}$ & $68.19(68.00)^{+0.62}_{-1.2}$ & $68.14(67.93)^{+0.65}_{-1.0}$\\
$n_\mathrm{s}              $ & $0.9688(0.9728)^{+0.0052}_{-0.0074}$ & $0.9915(0.98835)\pm 0.0046 $ & $0.9684(0.9705)^{+0.0055}_{-0.0075}$ & $0.9685(0.9704)^{+0.0053}_{-0.0076}$ & $0.9675(0.9665)^{+0.0049}_{-0.0063}$\\
$\Omega_\mathrm{b} h^2     $ & $0.02229(0.022368)^{+0.00016}_{-0.00019}$ & $0.02300(0.022933)\pm 0.00015$ & $0.02230(0.022351)^{+0.00016}_{-0.00021}$ & $0.02228(0.022233)^{+0.00016}_{-0.00020}$ & $0.02230(0.022264)^{+0.00016}_{-0.00020}$\\
$\Omega_\mathrm{c} h^2     $ & $0.1229(0.12402)^{+0.0019}_{-0.0033}$ & $0.1357(0.13449)^{+0.0017}_{-0.0020}$ & $0.1226(0.12386)^{+0.0019}_{-0.0035}$ & $0.1222(0.12143)^{+0.0015}_{-0.0032}$ & $0.1227(0.12219)^{+0.0017}_{-0.0035}$\\
$\log(10^{10} A_\mathrm{s})$ & $3.053(3.0533)\pm 0.014    $ & $3.090(3.0854)\pm 0.013    $ & $3.054(3.0565)\pm 0.015    $ & $3.052(3.0504)^{+0.014}_{-0.015}$ & $3.054(3.0521)^{+0.014}_{-0.015}$\\
$\tau_\mathrm{reio}        $ & $0.0582(0.0577)\pm 0.0061  $ & $0.0595(0.0581)\pm 0.0064  $ & $0.0577(0.0578)\pm 0.0062  $ & $0.0580(0.0579)\pm 0.0062  $ & $0.0578(0.0576)\pm 0.0061  $\\
\hline
$\Omega_\mathrm{m}         $ & $0.3128(0.3136)\pm 0.0074  $ & $0.3066(0.3068)\pm 0.0069  $ & $0.3135(0.3137)\pm 0.0072  $ & $0.3122(0.3121)\pm 0.0077  $ & $0.3138(0.3144)\pm 0.0073  $\\
$r_s h                     $ & $99.41(99.28)\pm 0.94      $ & $100.05(100.02)\pm 0.89    $ & $99.31(99.26)\pm 0.92      $ & $99.49(99.53)\pm 0.98      $ & $99.27(99.19)\pm 0.92      $\\
$S_8                       $ & $0.836(0.8404)\pm 0.015    $ & $0.877(0.8722)\pm 0.015    $ & $0.838(0.8433)\pm 0.016    $ & $0.834(0.8325)\pm 0.015    $ & $0.839(0.8382)\pm 0.016    $\\
\hline\hline
\end{tabular}
    \caption{The mean values and $1\sigma$ intervals (the best-fit values are in parentheses) of the parameters for the different EDE models with PR4 dataset.
    For one-tailed distributions, their $2\sigma$ bound is shown instead.}
    \label{tab:PR4}
\end{table*}

\begin{table*}[htb!]
\fontsize{6}{0}
    \centering
\begin{tabular} {| l | c| c| c| c| c|}
\hline\hline
 & axion-like EDE & $\phi^4$ AdS-EDE & canonical ADE & cold NEDE & $\phi^4$ Rock ‘n’ Roll\\
\hline
$f_\mathrm{EDE}(z_c)       $ & $< 0.0849(0.0660)          $ & $0.1156(0.1084)^{+0.0035}_{-0.0089}$ & $< 0.0835(0.0472)          $ & $< 0.107(0.081)            $ & $< 0.0614(0.0087)          $\\
$\log_{10}(z_c)            $ & $3.60(3.738)\pm 0.16       $ & $3.529(3.5202)^{+0.022}_{-0.026}$ & $3.521(3.446)^{+0.097}_{-0.21}$ & $3.70(3.591)^{+0.14}_{-0.31}$ & $3.56(3.571)^{+0.11}_{-0.15}$\\
$\Theta_\mathrm{ini}       $ & $2.29(2.929)^{+0.84}_{-0.085}$ &                              &                              &                              &                             \\
$3 \omega_\mathrm{NEDE}    $ &                              &                              &                              & $- \ (2.01)              $ &                             \\
\hline
$H_0                       $ & $69.11(69.89)^{+0.70}_{-1.0}$ & $72.65(72.36)^{+0.46}_{-0.54}$ & $68.88(69.20)^{+0.59}_{-0.99}$ & $69.13(70.29)^{+0.66}_{-1.2}$ & $68.84(68.41)^{+0.49}_{-0.94}$\\
$n_\mathrm{s}              $ & $0.9734(0.9795)^{+0.0053}_{-0.0075}$ & $0.9943(0.99134)\pm 0.0043 $ & $0.9721(0.9744)^{+0.0050}_{-0.0069}$ & $0.9734(0.9787)^{+0.0057}_{-0.0073}$ & $0.9712(0.9693)^{+0.0045}_{-0.0061}$\\
$\Omega_\mathrm{b} h^2     $ & $0.02238(0.022412)^{+0.00015}_{-0.00017}$ & $0.02308(0.023013)\pm 0.00015$ & $0.02240(0.022399)^{+0.00015}_{-0.00020}$ & $0.02238(0.022394)^{+0.00016}_{-0.00018}$ & $0.02240(0.022380)^{+0.00016}_{-0.00019}$\\
$\Omega_\mathrm{c} h^2     $ & $0.1218(0.12419)^{+0.0021}_{-0.0033}$ & $0.1343(0.13325)^{+0.0016}_{-0.0019}$ & $0.1211(0.12283)^{+0.0018}_{-0.0036}$ & $0.1216(0.12487)^{+0.0018}_{-0.0037}$ & $0.1211(0.11901)^{+0.0014}_{-0.0034}$\\
$\log(10^{10} A_\mathrm{s})$ & $3.055(3.0581)\pm 0.015    $ & $3.090(3.0849)\pm 0.014    $ & $3.055(3.0584)\pm 0.016    $ & $3.055(3.0598)\pm 0.015    $ & $3.054(3.0485)^{+0.014}_{-0.016}$\\
$\tau_\mathrm{reio}        $ & $0.0602(0.0598)^{+0.0060}_{-0.0067}$ & $0.0612(0.0597)\pm 0.0066  $ & $0.0597(0.0597)\pm 0.0061  $ & $0.0598(0.0594)\pm 0.0063  $ & $0.0598(0.0594)\pm 0.0061  $\\
\hline
$\Omega_\mathrm{m}         $ & $0.3034(0.3015)\pm 0.0054  $ & $0.2995(0.2996)\pm 0.0051  $ & $0.3039(0.3046)\pm 0.0052  $ & $0.3027(0.2994)\pm 0.0054  $ & $0.3041(0.30351)\pm 0.0050 $\\
$r_s h                     $ & $100.62(100.87)^{+0.66}_{-0.74}$ & $100.98(100.97)\pm 0.69    $ & $100.54(100.45)\pm 0.69    $ & $100.71(101.14)\pm 0.72    $ & $100.51(100.60)\pm 0.66    $\\
$S_8                       $ & $0.821(0.8243)\pm 0.012    $ & $0.863(0.8589)\pm 0.012    $ & $0.822(0.8290)^{+0.013}_{-0.016}$ & $0.821(0.8247)\pm 0.013    $ & $0.822(0.8137)^{+0.012}_{-0.015}$\\
\hline\hline
\end{tabular}
    \caption{The mean values and $1\sigma$ intervals (the best-fit values are in parentheses) of the parameters for the different EDE models with PR4 + DESI dataset.
    For one-tailed distributions, their $2\sigma$ bound is shown instead.}
    \label{tab:PR4_DESI}
\end{table*}

\begin{table*}[htb!]
\fontsize{6}{0}
    \centering
\begin{tabular} {| l | c| c| c| c| c|}
\hline\hline
 & axion-like EDE & $\phi^4$ AdS-EDE & canonical ADE & cold NEDE & $\phi^4$ Rock ‘n’ Roll\\
\hline
$f_\mathrm{EDE}(z_c)       $ & $< 0.0805(0.0633)          $ & $0.1157(0.1083)^{+0.0035}_{-0.0087}$ & $< 0.0818(0.0466)          $ & $< 0.101(0.073)            $ & $< 0.0605(0.0196)          $\\
$\log_{10}(z_c)            $ & $3.59(3.749)^{+0.16}_{-0.19}$ & $3.528(3.5259)^{+0.022}_{-0.025}$ & $3.525(3.461)^{+0.094}_{-0.22}$ & $3.70(3.611)^{+0.17}_{-0.32}$ & $3.55(3.502)^{+0.11}_{-0.15}$\\
$\Theta_\mathrm{ini}       $ & $> 0.376(2.93)             $ &                              &                              &                              &                             \\
$3 \omega_\mathrm{NEDE}    $ &                              &                              &                              & $- \ (2.02)              $ &                             \\
\hline
$H_0                       $ & $68.85(69.59)^{+0.63}_{-0.98}$ & $72.46(72.123)^{+0.46}_{-0.53}$ & $68.69(68.99)^{+0.55}_{-0.98}$ & $68.84(69.84)^{+0.59}_{-1.1}$ & $68.68(68.45)^{+0.48}_{-0.91}$\\
$n_\mathrm{s}              $ & $0.9719(0.9775)^{+0.0051}_{-0.0071}$ & $0.9932(0.99113)\pm 0.0043 $ & $0.9711(0.9726)^{+0.0050}_{-0.0070}$ & $0.9720(0.9766)^{+0.0053}_{-0.0071}$ & $0.9703(0.9680)^{+0.0044}_{-0.0060}$\\
$\Omega_\mathrm{b} h^2     $ & $0.02235(0.022359)^{+0.00015}_{-0.00017}$ & $0.02304(0.022986)\pm 0.00015$ & $0.02237(0.022395)^{+0.00015}_{-0.00020}$ & $0.02235(0.022387)^{+0.00016}_{-0.00018}$ & $0.02238(0.022309)^{+0.00015}_{-0.00019}$\\
$\Omega_\mathrm{c} h^2     $ & $0.1219(0.12430)^{+0.0019}_{-0.0032}$ & $0.1349(0.13367)^{+0.0015}_{-0.0018}$ & $0.1214(0.12315)^{+0.0017}_{-0.0035}$ & $0.1216(0.12450)^{+0.0015}_{-0.0034}$ & $0.1214(0.12135)^{+0.0014}_{-0.0034}$\\
$\log(10^{10} A_\mathrm{s})$ & $3.054(3.0563)\pm 0.015    $ & $3.090(3.0870)\pm 0.014    $ & $3.055(3.0588)\pm 0.015    $ & $3.054(3.0587)\pm 0.015    $ & $3.054(3.0522)^{+0.014}_{-0.016}$\\
$\tau_\mathrm{reio}        $ & $0.0596(0.0597)\pm 0.0062  $ & $0.0605(0.0600)\pm 0.0064  $ & $0.0594(0.0584)\pm 0.0061  $ & $0.0593(0.0589)\pm 0.0063  $ & $0.0594(0.0586)\pm 0.0061  $\\
\hline
$\Omega_\mathrm{m}         $ & $0.3058(0.3042)\pm 0.0051  $ & $0.3020(0.30241)\pm 0.0049 $ & $0.3062(0.3071)\pm 0.0050  $ & $0.3052(0.3025)\pm 0.0052  $ & $0.3062(0.30802)\pm 0.0049 $\\
$r_s h                     $ & $100.31(100.51)\pm 0.66    $ & $100.64(100.60)\pm 0.65    $ & $100.25(100.11)\pm 0.66    $ & $100.39(100.72)\pm 0.68    $ & $100.25(100.01)\pm 0.64    $\\
$S_8                       $ & $0.825(0.8277)\pm 0.012    $ & $0.868(0.8650)\pm 0.012    $ & $0.826(0.8333)^{+0.013}_{-0.015}$ & $0.824(0.8280)\pm 0.012    $ & $0.825(0.8271)^{+0.012}_{-0.015}$\\
\hline\hline
\end{tabular}
    \caption{The mean values and $1\sigma$ intervals (the best-fit values are in parentheses) of the parameters for the different EDE models with PR4 + DESI + PantheonPlus dataset.
    For one-tailed distributions, their $2\sigma$ bound is shown instead.}
    \label{tab:PR4_DESI_PP}
\end{table*}

\begin{table*}[htb!]
\fontsize{6}{0}
    \centering
\begin{tabular} {| l | c| c| c| c| c|}
\hline\hline
 & axion-like EDE & $\phi^4$ AdS-EDE & canonical ADE & cold NEDE & $\phi^4$ Rock ‘n’ Roll\\
\hline
$f_\mathrm{EDE}(z_c)       $ & $0.101(0.1057)^{+0.025}_{-0.023}$ & $0.1170(0.1099)^{+0.0039}_{-0.0093}$ & $0.101(0.1047)\pm 0.025    $ & $0.125(0.1331)^{+0.029}_{-0.026}$ & $0.076(0.0800)^{+0.021}_{-0.019}$\\
$\log_{10}(z_c)            $ & $3.64(3.734)^{+0.11}_{-0.14}$ & $3.527(3.5191)\pm 0.023    $ & $3.488(3.4934)^{+0.049}_{-0.038}$ & $3.587(3.584)^{+0.061}_{-0.079}$ & $3.540(3.5397)^{+0.034}_{-0.039}$\\
$\Theta_\mathrm{ini}       $ & $2.70(2.894)^{+0.27}_{+0.058}$ &                              &                              &                              &                             \\
$3 \omega_\mathrm{NEDE}    $ &                              &                              &                              & $2.12(2.028)^{+0.15}_{-0.22}$ &                             \\
\hline
$H_0                       $ & $71.25(71.35)\pm 0.89      $ & $72.33(72.03)\pm 0.55      $ & $70.77(70.84)^{+0.72}_{-0.81}$ & $71.65(72.09)\pm 0.98      $ & $70.85(70.94)\pm 0.83      $\\
$n_\mathrm{s}              $ & $0.9860(0.9879)\pm 0.0064  $ & $0.9925(0.98918)\pm 0.0045 $ & $0.9852(0.9855)\pm 0.0061  $ & $0.9861(0.9872)\pm 0.0063  $ & $0.9821(0.9823)\pm 0.0056  $\\
$\Omega_\mathrm{b} h^2     $ & $0.02258(0.022579)\pm 0.00019$ & $0.02303(0.022955)\pm 0.00015$ & $0.02271(0.022717)\pm 0.00019$ & $0.02257(0.022495)\pm 0.00021$ & $0.02272(0.022739)\pm 0.00018$\\
$\Omega_\mathrm{c} h^2     $ & $0.1282(0.12783)\pm 0.0031 $ & $0.1356(0.13452)^{+0.0017}_{-0.0021}$ & $0.1285(0.12889)\pm 0.0032 $ & $0.1288(0.12958)\pm 0.0031 $ & $0.1292(0.12995)\pm 0.0036 $\\
$\log(10^{10} A_\mathrm{s})$ & $3.068(3.0656)\pm 0.015    $ & $3.091(3.0870)\pm 0.013    $ & $3.075(3.0756)\pm 0.015    $ & $3.070(3.0701)\pm 0.014    $ & $3.073(3.0739)\pm 0.015    $\\
$\tau_\mathrm{reio}        $ & $0.0613(0.0614)\pm 0.0065  $ & $0.0600(0.0590)\pm 0.0063  $ & $0.0609(0.0602)\pm 0.0061  $ & $0.0608(0.0602)\pm 0.0063  $ & $0.0600(0.0593)\pm 0.0063  $\\
\hline
$\Omega_\mathrm{m}         $ & $0.2984(0.2967)^{+0.0062}_{-0.0071}$ & $0.3045(0.3048)\pm 0.0061  $ & $0.3032(0.3034)\pm 0.0063  $ & $0.2962(0.2938)\pm 0.0069  $ & $0.3040(0.3047)\pm 0.0063  $\\
$r_s h                     $ & $101.24(101.47)\pm 0.92    $ & $100.32(100.29)\pm 0.81    $ & $100.57(100.54)\pm 0.84    $ & $101.54(101.87)\pm 0.95    $ & $100.45(100.33)\pm 0.83    $\\
$S_8                       $ & $0.831(0.8266)^{+0.014}_{-0.017}$ & $0.874(0.8701)\pm 0.014    $ & $0.847(0.8488)\pm 0.017    $ & $0.833(0.8305)\pm 0.016    $ & $0.848(0.8512)\pm 0.018    $\\
\hline\hline
\end{tabular}
    \caption{The mean values and $1\sigma$ intervals (the best-fit values are in parentheses) of the parameters for the different EDE models with PR4 + SH0ES dataset.}
    \label{tab:PR4_SH0ES}
\end{table*}

\begin{table*}[htb!]
\fontsize{6}{0}
    \centering
\begin{tabular} {| l | c| c| c| c| c|}
\hline\hline
 & axion-like EDE & $\phi^4$ AdS-EDE & canonical ADE & cold NEDE & $\phi^4$ Rock ‘n’ Roll\\
\hline
$f_\mathrm{EDE}(z_c)       $ & $0.100(0.1128)^{+0.026}_{-0.023}$ & $0.1168(0.1085)^{+0.0038}_{-0.0093}$ & $0.099(0.1037)\pm 0.024    $ & $0.123(0.1304)^{+0.029}_{-0.026}$ & $0.073(0.0749)^{+0.021}_{-0.019}$\\
$\log_{10}(z_c)            $ & $3.64(3.586)^{+0.11}_{-0.13}$ & $3.527(3.5225)\pm 0.023    $ & $3.482(3.4805)^{+0.048}_{-0.040}$ & $3.586(3.599)^{+0.066}_{-0.076}$ & $3.540(3.5376)^{+0.033}_{-0.043}$\\
$\Theta_\mathrm{ini}       $ & $2.74(2.831)^{+0.23}_{+0.031}$ &                              &                              &                              &                             \\
$3 \omega_\mathrm{NEDE}    $ &                              &                              &                              & $2.13(2.058)^{+0.15}_{-0.22}$ &                             \\
\hline
$H_0                       $ & $71.20(71.60)\pm 0.82      $ & $72.56(72.249)^{+0.43}_{-0.50}$ & $70.89(71.05)^{+0.68}_{-0.77}$ & $71.52(71.74)\pm 0.85      $ & $70.94(71.02)\pm 0.81      $\\
$n_\mathrm{s}              $ & $0.9857(0.9868)\pm 0.0062  $ & $0.9936(0.99117)\pm 0.0042 $ & $0.9855(0.9859)\pm 0.0061  $ & $0.9856(0.9864)\pm 0.0061  $ & $0.9825(0.9827)\pm 0.0055  $\\
$\Omega_\mathrm{b} h^2     $ & $0.02256(0.022493)\pm 0.00019$ & $0.02306(0.023014)\pm 0.00015$ & $0.02270(0.022719)\pm 0.00019$ & $0.02256(0.022526)\pm 0.00021$ & $0.02273(0.022736)\pm 0.00018$\\
$\Omega_\mathrm{c} h^2     $ & $0.1280(0.12971)\pm 0.0030 $ & $0.1349(0.13355)^{+0.0015}_{-0.0019}$ & $0.1279(0.12843)\pm 0.0030 $ & $0.1288(0.12962)\pm 0.0030 $ & $0.1283(0.12846)^{+0.0036}_{-0.0032}$\\
$\log(10^{10} A_\mathrm{s})$ & $3.067(3.0709)\pm 0.014    $ & $3.091(3.0876)\pm 0.013    $ & $3.074(3.0746)\pm 0.015    $ & $3.070(3.0684)\pm 0.014    $ & $3.072(3.0727)\pm 0.015    $\\
$\tau_\mathrm{reio}        $ & $0.0613(0.0605)\pm 0.0064  $ & $0.0608(0.0600)\pm 0.0063  $ & $0.0610(0.0603)\pm 0.0060  $ & $0.0607(0.0597)\pm 0.0062  $ & $0.0606(0.0603)\pm 0.0063  $\\
\hline
$\Omega_\mathrm{m}         $ & $0.2983(0.29813)\pm 0.0047 $ & $0.3014(0.30117)\pm 0.0046 $ & $0.3010(0.30070)\pm 0.0046 $ & $0.2972(0.29684)\pm 0.0048 $ & $0.3013(0.30105)\pm 0.0046 $\\
$r_s h                     $ & $101.25(101.28)\pm 0.64    $ & $100.73(100.77)\pm 0.61    $ & $100.88(100.90)\pm 0.62    $ & $101.39(101.45)\pm 0.66    $ & $100.80(100.83)\pm 0.61    $\\
$S_8                       $ & $0.830(0.8345)\pm 0.013    $ & $0.868(0.8630)\pm 0.011    $ & $0.842(0.8435)\pm 0.015    $ & $0.834(0.8345)\pm 0.013    $ & $0.842(0.8421)^{+0.016}_{-0.014}$\\
\hline\hline
\end{tabular}
    \caption{The mean values and $1\sigma$ intervals (the best-fit values are in parentheses) of the parameters for the different EDE models with PR4 + SH0ES + DESI + PantheonPlus dataset.}
    \label{tab:PR4_SH0ES_DESI_PP}
\end{table*}

\section{Conclusions}
\label{sec:conclusions}

The EDE model, as a promising solution to the Hubble tension, increases the physical CDM density $\omega_c = \Omega_c h^2$ in order to compensate for its suppression of the gravitational potential when fitting the CMB power spectrum.
However, this does not mean that the density of matter $\Omega_m$ (and hence $r_s h$) can remain constant.
Analysis of Planck PR4 data in this work suggests that the EDE model actually prefers a lower $\Omega_m$ and a higher $r_s h$.
Hence, the consistency of EDE with the measurements of BAO and SNeIa in past analyses may in fact be a coincidence or due to large uncertainties.

On the other hand, there are some inconsistencies between recent BAO and SNeIa observations, and they also seem to deviate from the Planck results.
In this work, I focus on the recent DESI BAO results.
It prefers smaller $\Omega_m$ and larger $r_s h$ relative to the Planck 2018 results, and they are consistent with the direction in which the EDE shifted the parameters when fitting the CMB observations.
Therefore DESI BAO will prefer the EDE model.
For most of the EDE models tested in this work, the DESI BAO measurement does slightly enhance the preference for EDE models, and raises $H_0$ for all of the EDE models here, although it is not sufficiently strong to be compatible with the SH0ES results.
Meanwhile, thanks to the lower $\Omega_m$, the $S_8$ tension is also slightly mitigated.

Pantheon+ measurements for SNeIa tend to deviate $\Omega_m$ in the opposite direction from DESI.
However, in this work, it is found that the trend of parameter shifts in the joint analysis is still dominated by DESI.
And this drive from DESI BAO is unforeseen in past SDSS results.
Additionally, in the joint analysis with measurements of the local universe from SH0ES, the preferences for EDE and the constraints on $H_0$ are dominated by SH0ES.

Throughout this work, I assume the late universe remains $\Lambda$CDM.
However, early new physics alone may not be sufficient to fully resolve the Hubble tension (see e.g.~\cite{Vagnozzi:2023nrq} for a review), and thus modifications to the late and/or local universe may be required.
The tension with large-scale structures also calls for modifications for the late universe.
At the same time, early dark energy models do not reconcile the tension between BAO and SNeIa regarding $\Omega_m$, which also suggests potential unknown systematic errors or modifications for the late or local universe.
Actually, the recent DESI Y1 BAO might be a hint for some modifications on the late-time expansion history when combined with SNeIa and CMB~\cite{DESI:2024mwx}.
The constraints on early dark energy in these cases will require further study (see e.g.~\cite{Wang:2024dka}).

Furthermore, in this work I focus only on the Planck satellite observations of the CMB, while it is known that some recent ground-based CMB observations seem to prefer the EDE model (e.g.~\cite{Jiang:2021bab,Hill:2021yec,Poulin:2021bjr,Jiang:2022uyg,LaPosta:2021pgm,Smith:2022hwi}).
It would be interesting to examine the results found in this paper in the context of these CMB observations.

In summary, this work emphasizes the importance of $\Omega_m$ and $r_s H_0$ measurements including BAO and SNeIa in constraining the EDE model.
The upcoming DESI and Euclid BAO data releases will help clarify the findings here.
Meanwhile, other measurements for $\Omega_m$ or even model-independent measurements will be helpful.

\begin{acknowledgments}
\noindent I would like to thank Sunny Vagnozzi and Yun-Song Piao for their discussions. I acknowledge the use of high performance computing services provided by the International Centre for Theoretical Physics Asia-Pacific cluster and Scientific Computing Center of University of Chinese Academy of Sciences.
J.-Q.J.\ acknowledges support from the Joint PhD Training program of the University of Chinese Academy of Sciences.
\end{acknowledgments}

\bibliography{edeafterdesi}
\end{document}